\documentclass[aps,twocolumn,a4paper,showpacs]{revtex4}
\usepackage[colorlinks=true, pdfstartview=FitV, linkcolor=blue, citecolor=red, urlcolor=magenta, breaklinks=true]{hyperref}
\usepackage{graphicx}
\usepackage[all]{xy}
\usepackage{amsmath}
\usepackage{amssymb}
\usepackage{color}
\usepackage{subeqnarray}
\usepackage{subfigure}
\usepackage{epstopdf}

\newcommand{\bb}{\bibitem}
\newcommand{\bes}{\begin{subequations}}
\newcommand{\ees}{\end{subequations}}
\def\ben{\begin{eqnarray}}
\def\een{\end{eqnarray}}
\newcommand{\bens}{\begin{subeqnarray}}
\newcommand{\eens}{\end{subeqnarray}}
\def\be{\begin{equation}}
\def\ee{\end{equation}}
\def\tanh{\text{tanh}}
\def\cos{\text{cos}}
\def\sin{\text{sin}}
\def\sech{\text{sech}}
\def\sec{\text{sec}}
\def\ln{\text{ln}}
\def\e{\text{e}}
\begin{document}
\title{High Temperature Effects on Compactlike Structures} 
\author{D. Bazeia} 
\author{E.E.M. Lima}
\author{L. Losano}  
\affiliation{Departamento de F\'\i sica Universidade Federal da Para\'iba, 58051-900 Jo\~ao Pessoa PB, Brazil}
\pacs{11.10. Lm, 11.27. +d}
\begin{abstract}
In this work we investigate the transition from kinks to compactons at high temperatures. We deal with a family of models, described by a real scalar field with standard kinematics, controlled by a single parameter, real and positive. The family of models supports kinklike solutions, and the solutions tend to become compact when the parameter increases to larger and larger values. We study the one-loop corrections  at finite temperature, to see how the thermal effects add to the effective potential. The results suggest that the symmetry is restored at very high temperatures.
\end{abstract}
\maketitle

\section{Introduction}

Topological defects are of current interest and have attracted a great deal of attention in high energy physics \cite{r1,VS,MS} and in other areas of nonlinear science; see, e.g., Ref.~\cite{DM}. Among them, the simplest topological structures are kinks, which appear in models described by real scalar fields in $(1,1)$ spacetime dimensions. They can be embedded in $(3,1)$ space-time dimensions as domain walls, and usually evolve under standard kinematics, subjected to a potential that develops spontaneous symmetry breaking.

Under specific conditions, another kind of kinklike defect, called compacton, appears in models with generalized kinematics that include a nonlinear dispersion \cite{RH,K2,BLMO,CoA1}. These structures are nontrivial configurations with compact support, and have been studied in distinct contexts in \cite{comp1,comp2,comp3,GAGA,MA}. In particular, in Ref.~\cite{MA} one shows how a kinklike solution can be transformed into a compact structure, driven by a single parameter, even though the model engenders standard kinematics.

Motivated by this fact, in the current work we are interested to study how the compact structure is impacted by the presence of quantum corrections. The idea was advanced in \cite{BV}, where the one-loop shift of the energy of a compacton was calculated in a model with modified kinematics. Here, however, we study a model first introduced in \cite{MA}, with standard kinematics. Due to the standard kinematics, we could calculate the one-loop correction to get to the effective potential following the usual route. In this sense, the present investigation is indirect, since we will study the effective potential instead of the effective action. However, if the thermal effects are able to restore the symmetry at some critical temperature $T_c$, the system cannot support defect structure anymore, when the temperature is higher or equal to the critical one.

As one knows, in the standard scenario \cite{Jac}, the high temperature effects allows the symmetry restoration, leading to a phase transition where topological structures appear below the critical temperature. Although this is the general wisdom, there are models where the symmetry is never restored \cite{MO,Drac}, implying the absence of phase transition at high temperatures. So, it is of current interest to study how the thermal effects act to control the smooth transition that changes the kink into a compact structure. 

To ease the investigation, we organize the work as follows. In the next section, we briefly review models of a single real scalar field and in Sec.~\ref{sec-2} we describe the connection between kinks and compactons. In Sec.~\ref{sec-3} we develop the procedure to obtain the effective potential in the high temperature limit, and we study how it behaves in the compact limit. As far as we can see, the present calculations provide the first results to suggest how compact kinks behave at finite temperature. We end the work including our comments and conclusions in Sec.\ref{sec-com}.

\section{Generalities}\label{sec-1}

We start our investigation from the general Lagrange density
describing a relativistic system driven by a single real scalar field defined as 
\be
\label{lagran}
{\mathcal L}=\frac{1}{2}\partial_\mu \phi \partial^\mu \phi-V(\phi),
\ee
where $V(\phi)$ is a potential which specifies self-interactions of the scalar field. Our notation is usual: we consider natural units $\hbar=c=1$, and work in the four-dimensional spacetime with the metric $(+,-,-,-)$. In this case, the scalar field has dimension of energy, and the spacetime coordinates have dimension inverse of energy. For simplicity, we redefine the field and the spacetime coordinates to make them dimensionless. We now suppose that $\phi=\phi(x,t)$ to find the following equation of motion
\ben
\label{eom}
\frac{\partial^2 \phi}{\partial t^2}- \frac{\partial^2 \phi}{\partial x^2}+\frac{d V}{d\phi}&=&0.
\een
We search for defect structures, considering potentials that can be written as $V(\phi)=(1/2){W_\phi}^2$, where  $W_\phi=d W/ d\phi$ and $W = W(\phi)$ is a smooth function of the scalar field. We suppose that the scalar field is static, $\phi = \phi(x)$, and so the equation of motion becomes
\be
\frac{d^2 \phi}{d x^2}=W_\phi W_{\phi\phi}.
\ee
This equation can be reduced to the first-order equation 
\be\label{foe}
\frac{d \phi}{dx}=W_{\phi}
\ee
In fact, there are two equations, and this is controlled by the sign of $W$, which does not modify the theory, but changes the first-order equation. Each equation has one solution, and the solution is such that $\phi (x \to \infty)$ and $\phi (x \to -\infty)$ are neighbor minima of the potential; also, the derivative of the solution vanishes asymptotically. 

We can also check that each pair of neighbors minima ${\bar\phi}_a$ and ${\bar\phi}_b$ of the potential describes a topological sector with minimum energy given by $E=|\Delta W|$, where $\Delta W=W(\bar{\phi}_{a})-W(\bar{\phi}_b)$. Thus, we may get the energy of the solution even without knowing the solution explicitly.

Another important feature to be studied in order to evaluate quantum effects is the linear stability of the solutions, which exposes the behavior of the static field when submitted to small fluctuations. Consider a perturbation of the type $\phi(x,t)= \phi(x) +\eta_{n}(x)\cos(\omega_{n} t)$, where $\phi(x)$ is the static solution. Substituting this into (\ref{eom}) and expanding up to first-order in $\eta(x,t)$ we get a Schr\"odinger-like equation $H\eta_{n}(x)=\omega_{n}^{2}\eta_{n}(x)$, such that
\be
\left(-\frac{d^2}{dx^2}+U(x)\right)\eta_{n}(x)=\omega_{n}^{2}\eta_{n}(x),
\ee
where 
\be
U(x)=W_{\phi\phi}^2+W_\phi W_{\phi\phi\phi}
\ee
has to be calculated at the static solution, $\phi={\phi(x)}$. We can also write the Hamiltonian as $H=A^\dag A$, in terms of the operators
\be
A^\dag=-\frac{d}{dx} - W_{\phi\phi}\,\,\,\,\, \mbox{and} \,\,\,\,\,
A=\frac{d}{dx} -W_{\phi\phi}. \nonumber
\ee
This shows that $H$ is non negative. Thus, the solution of the first-order equation \eqref{foe} is classically or linearly stable.

\section{From kinks to compactons}
\label{sec-2}

The very important model that exhibit spontaneous symmetry breaking is the $\phi^4$ theory. The potential can be written as
\be
\label{phi4}
V(\phi)=\frac{1}{2}(1-\phi^2)^2,
\ee
where we are using dimensionless units for the field and the space and time coordinates, as already informed. This model has minima $\bar{\phi}_\pm=\pm 1$ and mass $m^2=4$. The topological solution connecting the two minima is the kinklike solution given by $\phi(x)=\tanh(x)$, with energy $E=4/3$ and energy density of the form
\be
\label{rho1}
\rho(x)=\sech^4(x).
\ee
The stability potential results in the well-known modified Poeschl-Teller potential \cite{PT}
\be
U(x)=4-6\,\sech^2(x).
\ee
This potential has the zero mode with energy $E_{0} = 0$, and one excited state with $E_1=3$. Also, there is an uncountable number of (non localized) states with energy $E\geq4$.   

An alternative to obtain compact kinks through the scalar field model involves a change in the kinematic term of Lagrange density \cite{BLMO}
\be
{\mathcal L}=-\frac{1}{4}\left(\partial_\mu \phi \partial^\mu \phi\right)^2-\frac{3}{2}V(\phi),
\ee 
with $V(\phi)$ given by Eq.~\eqref{phi4}. Here the equation of motion for static field is affected by a new nonlinear factor; it becomes
\be
\left(\frac{d\phi}{dx}\right)^2\frac{d^2\phi}{dx^2}=-\phi(1-\phi^2).
\ee
For this model we have the compact-like solution $\phi(x)=\sin(x)$ for $-\pi/2\leq x \leq \pi/2$, $\phi(x)=-1$ for $x < -\pi/2$, and $\phi(x)=1$ for $x > \pi/2$. The energy density is $\rho(x)=\cos^4(x)$ in the interval $-\pi/2\leq x \leq \pi/2$ and it vanishes outside this compact interval. By integration of $\rho(x)$ in all space we get the energy $E=3\pi/8$. In this situation, the  compact kink is stable with stability potential given by 
\be
U(x)=-4+2\,\sec^2\left(x\right)
\ee
for $-\pi/2\leq x \leq \pi/2$, and it is infinite for $|x|>\pi/2$. Here, all the states are bound states, and the corresponding energies are $E_{n}=\sqrt{n(n+4)}$, for $n=0,1,2,\cdots$. We then see that standard kinks and compact kinks appear from distinct physical contexts, so it is of current interest to further examine how the compact structures behave beyond the classical level. 

\begin{figure}
\includegraphics[width=7cm]{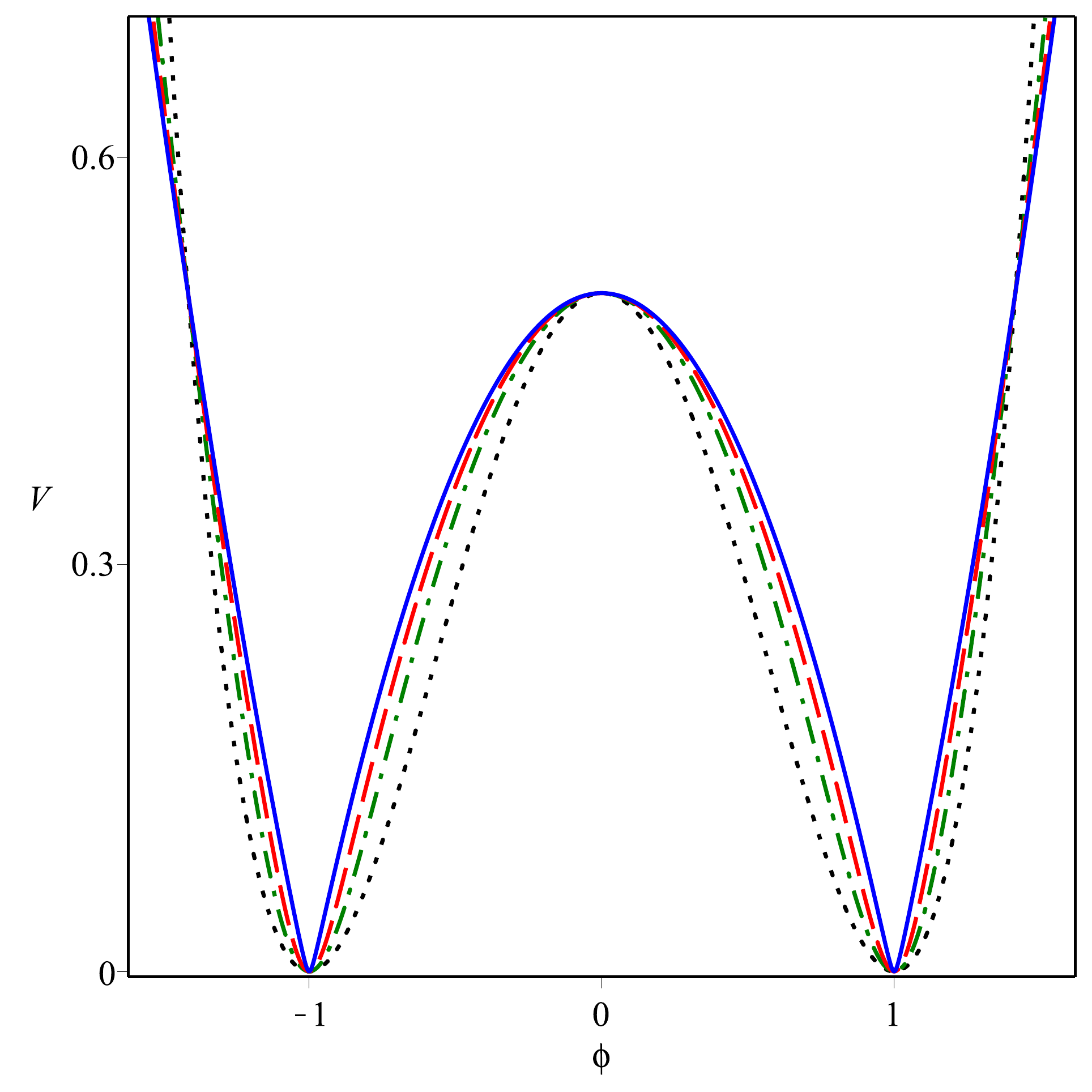}
\caption{The Potential described by Eq.~(\ref{potta}), depicted with the black (dotted), green (dotted-dashed), red (dashed), and blue (solid) lines for $\alpha=0,2,6,40$, respectively, showing how it behaves as one increases $\alpha$.}
\label{fig1}
\end{figure}

Other studies revealed other possible routes to find compact solutions; see, e.g., Refs.~\cite{comp3,GAGA,MA}. In particular, in Ref.~\cite{MA} one developed a class of models that smoothly convert kinks into compactons using standard kinematics. This fact motivated us to further investigate the model bellow, with the potential
\be
\label{potta}
V_{\alpha}(\phi)=\frac{1}{2\alpha}\left(\sqrt{1+4\alpha\left(1+\frac{\alpha}{2}\right)V(\phi)}-1\right),
\ee
with $V(\phi)$ given by \eqref{phi4}. This is a non-negative potential including two degenerate ground states $\bar{\phi}_\pm=\pm 1$. Here $\alpha$ is a positive real parameter that control the transformation of kinks into compactons. We illustrate this model depicting the potential \eqref{potta} in Fig.~\ref{fig1} for some values on $\alpha$. We also used $\alpha=100$ and $1000$, but they gives practically the same curve we have for $\alpha=40$, so we omitted them there. 

\begin{figure}
\includegraphics[width=7cm]{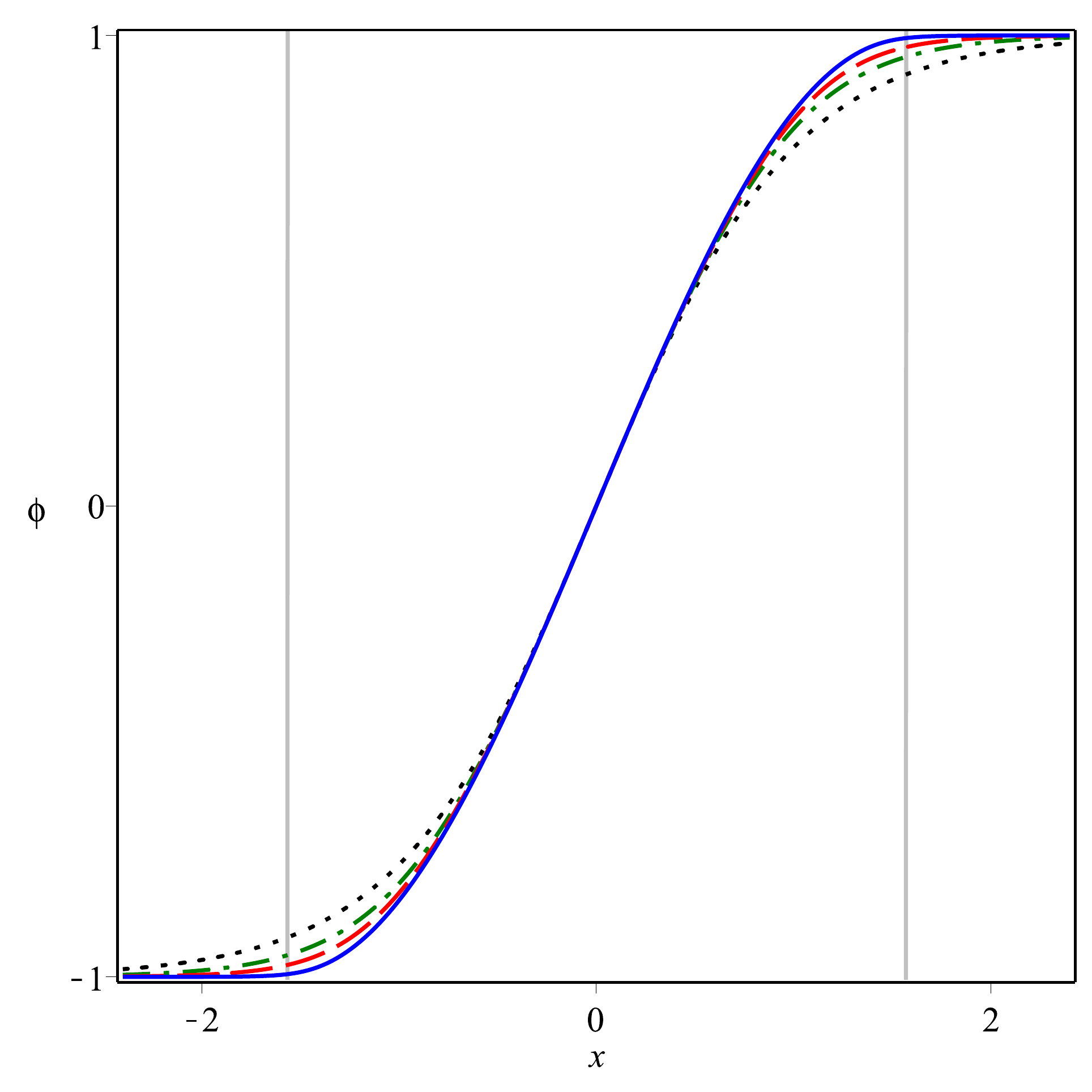}
\caption{The kinklike solution $\phi(x)$, depicted as in Fig.~\ref{fig1}, showing how the solution tends to become compact as $\alpha$ increases to larger and larger values.}
\label{fig2}
\end{figure}

The equation of motion is
\be
\frac{d^2\phi}{dx^2}=\frac{1+\alpha/2}{\sqrt{1+4\alpha\left(1+\frac{\alpha}{2}\right)V(\phi)}}\frac{dV}{d\phi}.
\ee
This equation is solved numerically and solutions are depicted in Fig.~\ref{fig2} for several values of
$\alpha$, with the two vertical gray lines showing the values $x=\pm\pi/2$. It is remarkable that the solution tends to become compact as $\alpha$ increases to larger and larger values. The second derivative of the potential \eqref{potta} is
\ben
\frac{d^2V}{d\phi^2}&=&\frac{1+\alpha/2}{[1+4\alpha\left(1+\frac{\alpha}{2}\right)V(\phi)]^{1/2}}\frac{d^2V}{d\phi^2} \nonumber \\ 
& &- \frac{2\alpha(1+\alpha/2)^2}{[1+4\alpha\left(1+\frac{\alpha}{2}\right)V(\phi)]^{3/2}}\left(\frac{dV}{d\phi}\right)^2.
\een
It gives the (squared) mass at the two minima, $m_{\alpha}^{2}=4+2\alpha$, which increases linearly with $\alpha$. We recall from \cite{MA} that one needs that the mass increases to larger and larger values to make the solution compact. Also, for $\alpha$ very large $(1/\alpha\ll1)$ we can expand the potential \eqref{potta} to get

\be\label{alphal}
V_\alpha(\phi)=\sqrt{V(\phi)/2}+\frac{1}{2\alpha}\left(\sqrt{2V(\phi)}-1\right)+O(\alpha^{-2}).
\ee
As a consequence, in the limit $1/\alpha \rightarrow 0$ the potential becomes
\be
V_c(\alpha)=\frac{1}{2}|1-\phi^2|. 
\ee
The kink described by $V_c(\phi)$ has a compact behavior, that is $\phi(x)=\sin(x)$ for $|x| \leq \pi/2$, $\phi(x)=-1$ for $x <-\pi/2$ and $\phi(x)=1$ for $x > \pi/2$. The energy density has the form $\rho_c(x)=\cos^2(x)$ when $|x| \leq \pi/2$, and it vanishes outside this compact interval. So we get the energy $E_c=\pi/2$.

In contrast, for small values of $\alpha$ the potential becomes
\be\label{alphas}
V_\alpha(\phi)=V(\phi)+\frac{\alpha}{2}V(\phi)\left(1-2V(\phi)\right)+O(\alpha^2).
\ee
In the limit $\alpha\rightarrow 0$  we get back to $\phi^4$ model \eqref{phi4} which supports the standard kink. 
 
Although the above results are known, they are important and will guide us toward the next investigation, where we concentrate on the effective potential at finite temperature.

\section{Thermal effects}\label{sec-3}

Let us now examine the effective potential, that is, the thermal effects at one-loop level for the model under investigation, described by the classical potential \eqref{potta}. We follow the standard route \cite{Jac}. We firstly note from \eqref{alphal} and \eqref{alphas} that the one-loop corrections make sense for $\alpha=0$, and also, for larger and larger values; so, in this section we investigate the thermal effects for $\alpha=0$ and for $\alpha\geq20$.

Toward this goal, it is convenient to introduce the generating functional 
\be
Z[J]=\frac{\int{D\phi\e^{iS[\phi]+i\int{J(x)\phi(x)d^4x}}}}{\int{D\phi\e^{iS[\phi]}}},
\ee 
with a source term $J$ and the action $S[\phi]=\int{d^4x{\mathcal L}}$. This gives the $n$-point Green functions written as
\be
g^{(n)}(x_1,...,x_n)=\frac{1}{i^n}\left.\frac{\delta^{n}Z[J]}{\delta J(x_1)...\delta J(x_n)}\right|_{J(x)=0}.
\ee 
For our objective only connected Green's functions are important, and they can be derived from the functional $W[J]=-i\ln Z[J]$. To examine quantum effects we must work in terms of the effective action, the generating functional of one particle irreducible Green's functions, $\Gamma[\phi_c]$, defined as
\ben
\Gamma[\phi_c]&=&W[J]-\int{d^4x J(x)\phi_c(x)},
\een
with $\phi_c(x)=\delta W[J]/\delta J(x)$ known as classical field that, when evaluated at $J=0$, results in the vacuum expectation value of the field $\bar{\phi}$ which engenders translational invariance. These conditions allow us to introduce the effective potential through
\be
\Gamma[\bar{\phi}]=-\Omega V_{eff}(\bar{\phi}),
\ee
where $\Omega$ is the spacetime volume.

The correction to the effective potential at one-loop level in Euclidean momentum space is given by
\be
V_{eff}(\phi)=V_{\alpha}(\phi)+\frac{1}{2}\int{\frac{d^4k}{(2\pi)^4}\ln[k^2+V_{\alpha}''(\phi)]},
\ee
where $V_{\alpha}''(\phi)=d^2V_{\alpha}/d\phi^2$; here and below we omit the bar in the scalar field, for simplicity. Finite temperature effects are introduced as in 
Ref.~\cite{Jac}, so the one-loop approximation gives
\be
V_1(\phi)=\frac{1}{2\beta}\int{\frac{d^3k}{(2\pi)^3}\sum_{n=-\infty}^{\infty}{\ln\left[\left(\frac{2\pi n}{\beta}\right)^2 +\vec{k}^2+V_{\alpha}''(\phi)\right]}}.
\ee
We implement the sum to obtain the finite temperature contribution
\be
V_1^\beta(\phi)=\frac{1}{2\pi^2\beta}\int{dk k^2 \ln\left(1-\e^{-\beta\sqrt{k^2 + V_{\alpha}''(\phi)}}\right)}.
\ee
The integration above can be achieved approximately, in the high temperature limit $T\gg m$. In this case, the effective potential becomes
\be
\label{effP}
V_{eff}(\phi)=V_{\alpha}(\phi)+\frac{T^2}{24}V_{\alpha}''(\phi).
\ee
Rewriting the classical potential (\ref{potta}) as
\be
V_\alpha(\phi)=\frac{\lambda}{2\alpha}\left(\sqrt{1+4\alpha\left(1+\frac{\alpha}{2}\right)V(\phi)}-1\right),
\ee
where $\lambda$ is a constant that indicates the strength of self-interactions, the mass is then modified and becomes $m_\alpha^2=\lambda(4+2\alpha)$, so the corresponding validity limit of the approximation is
$\lambda(4+2\alpha)/T^2\ll 1$. We use this and more, for clarity, instead of \eqref{effP} we modify the effective potential to write
\be\label{ueff}
U_{eff}(\phi)=V_{eff}(\phi)-V_{eff}(0).
\ee
It obeys $U_{eff}(0)=0$ and helps us better understand how the thermal effects contribute at high temperatures. 

\begin{figure}[t!]
\includegraphics[height=7cm,width=7cm]{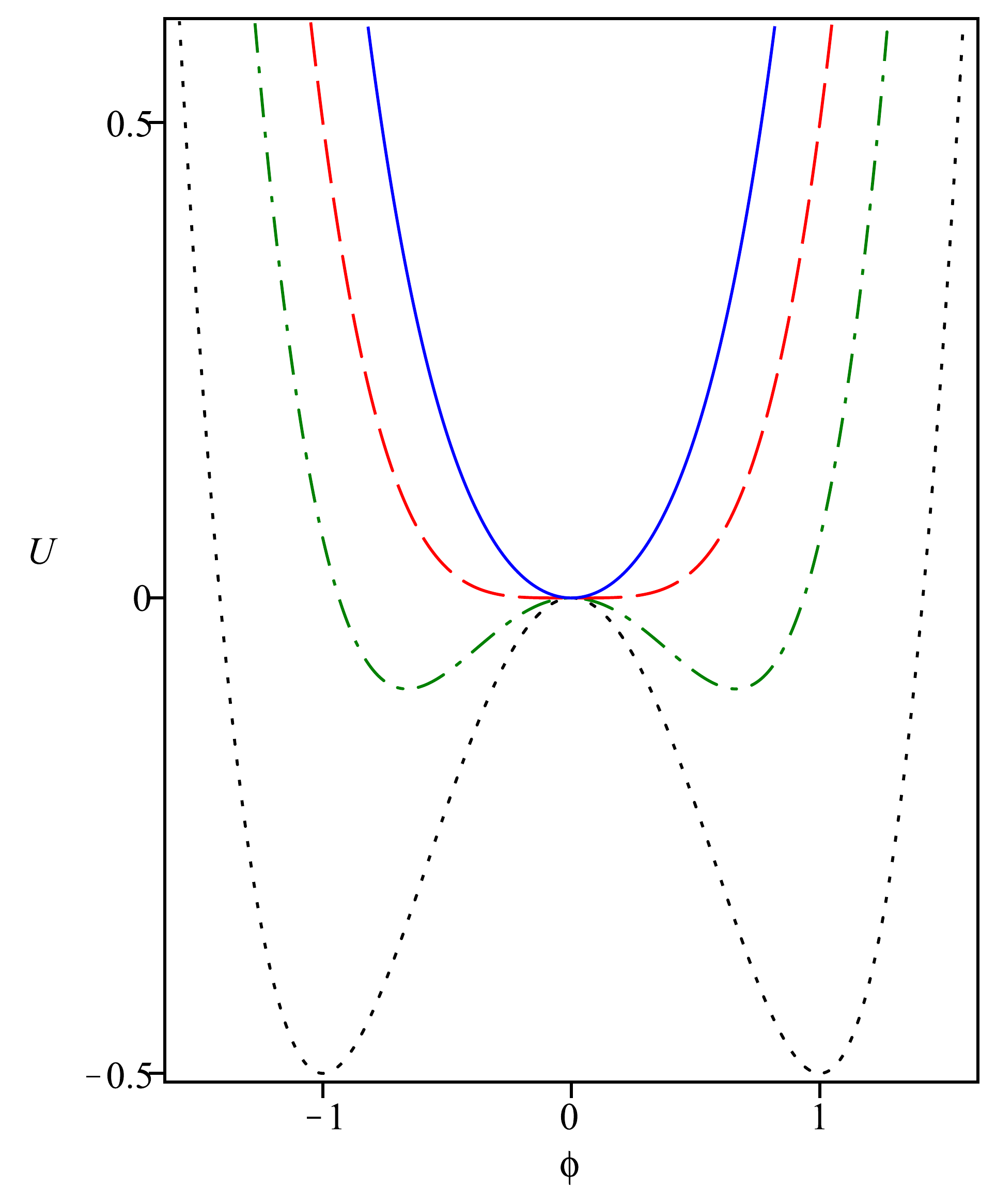}\;
\includegraphics[height=7cm,width=7cm]{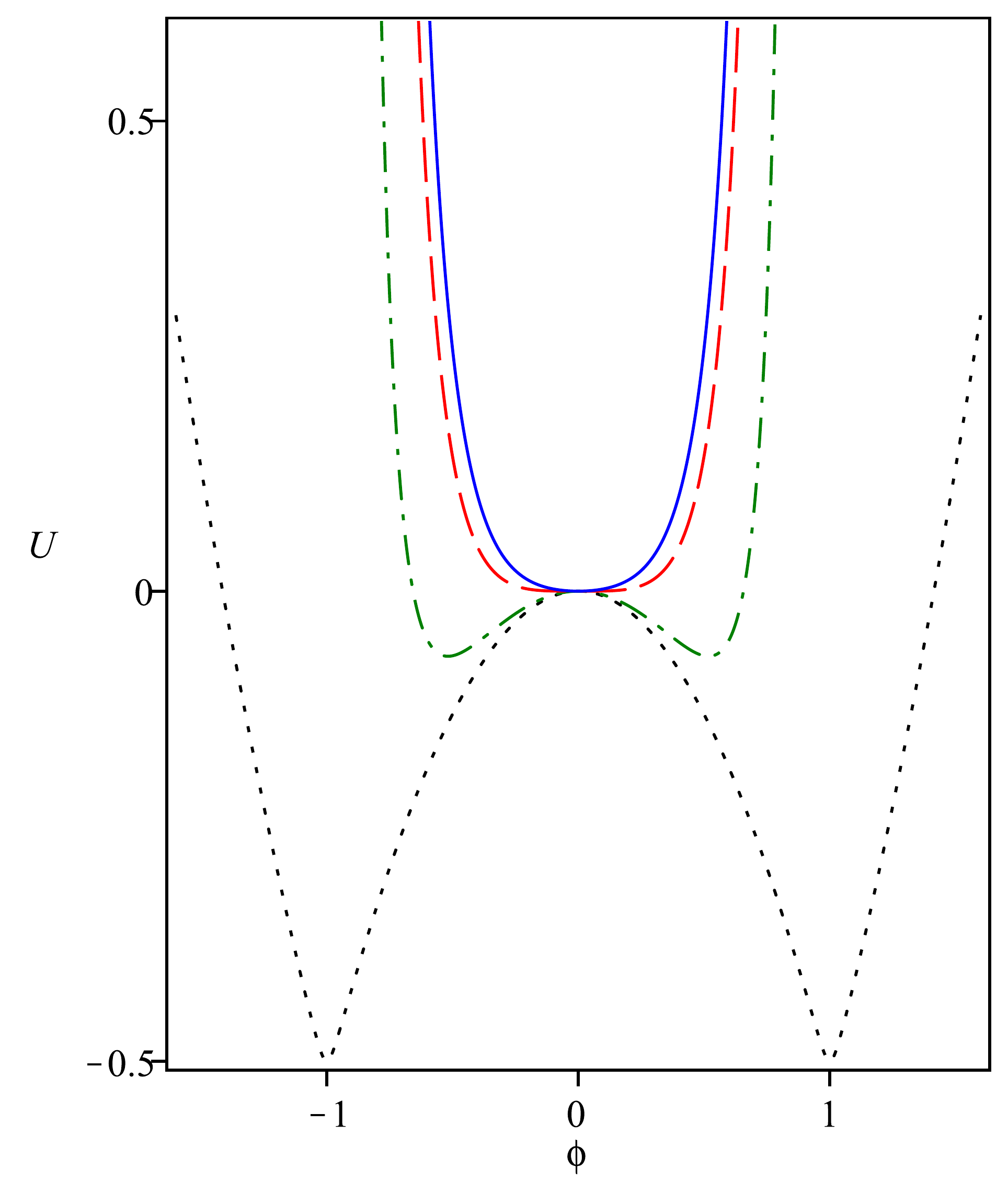}
\caption{The effective potential described by Eq.~\eqref{ueff}, depicted in the top panel for $\alpha=0$, showing the classical potential (dotted, black line), and the effective potential for $T=1.5$ (dotted-dashed, green line), $T=2.0$ (dashed, red line)  and $T=2.5$ (solid, blue line). In the bottom panel it is depicted for $\alpha=20$, for the classical potential (dotted, black line), and for $T=20$ (dotted-dashed, green line), $T=42$ (dashed, red line)  and $T=50$ (solid, blue line).}
\label{fig3}
\end{figure}

Here we are working in (3,1) spacetime dimensions. Hence, the potential may not be of interest for small values of $\alpha$, just because the theory can become nonrenormalizable, unless $\alpha=0$. On the other hand, we can work with large $\alpha$ in a manner consistent with the compact limit.

\begin{figure}[h!]
\;\;
\includegraphics[width=7cm]{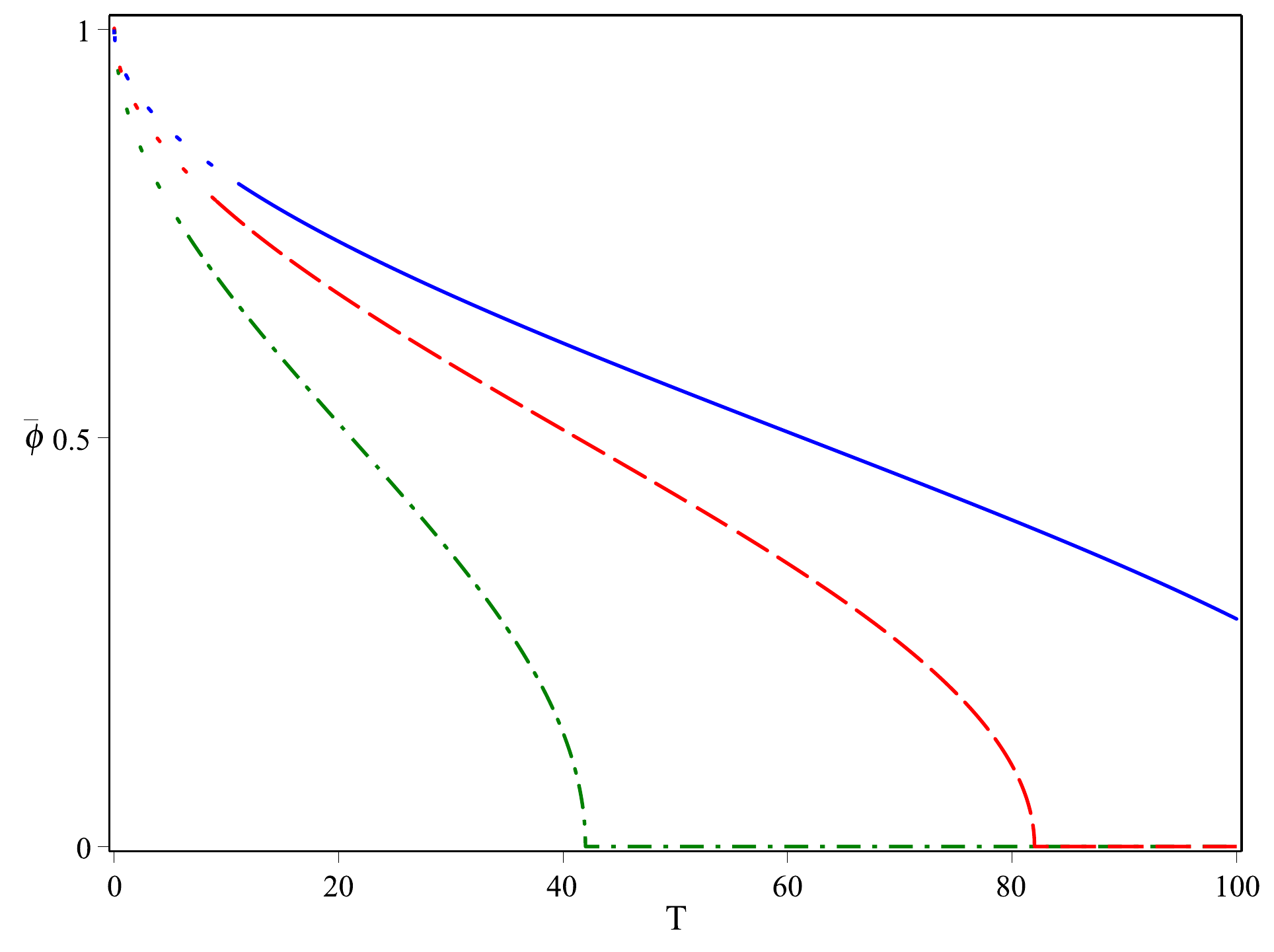}
\includegraphics[width=7.2cm]{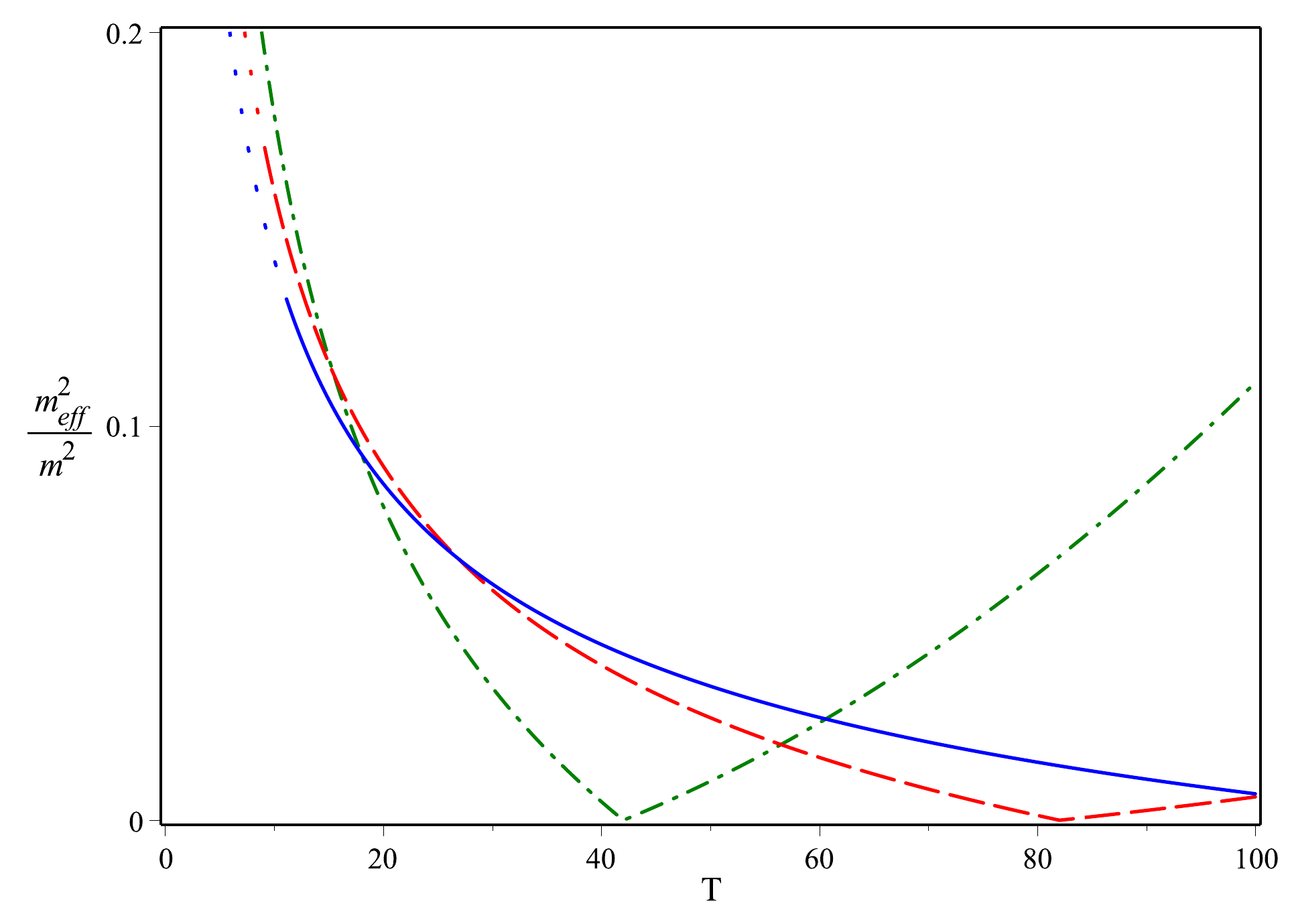}
\caption{The behavior of the minimum of the effective potential (top panel) and the effective mass \eqref{em} (bottom panel), depicted as a function of the temperature, with the dotted-dashed (green), dashed (red), and solid (blue) lines representing  the cases $\alpha=20,40$, and $60$, respectively.}\label{fig4}
\end{figure}

We want to explore how the finite temperature effects changes the symmetry breaking process for the model under consideration. Thus, in Fig.~\ref{fig3} we display the effective potential obtained in \eqref{ueff} for some values of $\alpha$ and $T$. These parameters should be chosen in a way that makes the approach consistent. The top panel shows, for $\alpha=0$, the well known phase transition that eliminate the spontaneous symmetry breaking, at a given critical temperature, in the $\phi^4$ theory. Otherwise, for $\alpha$ large enough, in the bottom panel one depicts the effective potential $U_{eff}(\phi)$ for several distinct values of $T$. The behavior suggests that the symmetry is restored at very high temperatures. This behavior can also be seen in Fig.~\ref{fig4}, where we depict the minimum of the potential as a function of $T$, and the effective (squared) mass, $m^2_{eff}/m^2$, where 
\be\label{em}
m^{2}_{eff}=\left.\frac{d^2U_{eff}}{d\phi^2}\right|_{\bar{\phi}}.
\ee
In Fig.~\ref{fig4}, the dotted-line segments indicates the region where the high temperature $(T\gg m)$ may not be reliable. From the behaviors depicted in Figs.~\ref{fig3} and \ref{fig4}, we note that the symmetry is restored at very high temperatures, and that the critical temperature $T_c$ increases as the parameter $\alpha$ increases. In fact, we have found a linear relation between $\alpha$ and the critical temperature; it is given by 
\be\label{tc}
T_c=2\alpha+2.
\ee
This result was obtained analytically, as follows: the behavior of the potential depicted in
Fig.~\ref{fig3} shows that the local maximum at $\phi=0$ changes to become a minimum as the temperature increases from values below the critical temperature to higher ones, above $T_c$. One then uses 
$U^{\prime\prime}_{eff}(\phi=0)=0$ to get to Eq.~\eqref{tc}. In the case $\alpha=0$ the critical temperature is $T_c=2$, and for $\alpha=20$ it is $T_c=42$, as shown in
Figs.~\ref{fig3} and \ref{fig4}.

\section{Comments and conclusions}\label{sec-com}

In this work, we investigated a model described by a real scalar field with standard kinematics. We considered a potential controlled by a single parameter, $\alpha$, which can be used to describe the smooth modification of defect solutions, changing kinks into compact structures. 

At the quantum level, we studied the thermal effects, investigating how the temperature-dependent one-loop corrections contribute to the effective potential. The results appear to be consistent for larger values of $\alpha$, and they suggest that the symmetry is restored at high temperatures, with the critical temperature $T_c(\alpha)$ presenting a linear relation with $\alpha$, which control the compact limit.

Strictly speaking, the compact structure appears in the limit $1/\alpha\to0$. However, the one-loop corrections to the effective potential seem to be reliable for larger values of $\alpha$, with the critical temperature $T_c=2\alpha+2$ leading to symmetry restoration at high temperature. In this sense, the results of the work suggest that although kinks and compact kinks appear from distinct physical contexts, their finite temperature effects act similarly, allowing for symmetry restoration at very high temperatures. 

\section*{Acknowledgments}

This work is partially supported by CNPq, Brazil. DB acknowledges support from grants 455931/2014-3 and 06614/2014-6, EEML acknowledges support from grant 160019/2013-3, and LL acknowledges support from grants 307111/2013-0 and 447643/2014-2.


\end{document}